\documentclass[12pt,a4paper]{iopart}

\usepackage{graphicx}
\usepackage{latexsym}
\usepackage{amsfonts}
\usepackage{amssymb}
\usepackage{bbm}
\usepackage{times}
\usepackage{color}

\newcommand{\be}{\begin{equation}}
\newcommand{\ee}{\end{equation}}
\newcommand{\bes}{\begin{equation*}}
\newcommand{\ees}{\end{equation*}}
\newcommand{\bel}[1]{\begin{equation}\label{#1}}
\newcommand{\bea}{\begin{eqnarray}}
\newcommand{\eea}{\end{eqnarray}}
\newcommand{\ba}{\begin{array}}
\newcommand{\ball}{\begin{array}{ll}}
\newcommand{\bacl}{\begin{array}{cl}}
\newcommand{\bacll}{\begin{array}{cll}}
\newcommand{\bal}{\begin{array}{l}}
\newcommand{\bac}{\begin{array}{c}}
\newcommand{\ea}{\end{array}}
\newcommand{\N}{{\mathbb{N}}}
\newcommand{\R}{{\mathbb{R}}}
\newcommand{\E}{{\mathbb{E}}}
\newcommand{\ebo}{{\mathbf{E}}}
\renewcommand{\P}{{\mathbb{P}}}

\newtheorem{thm}{Theorem}
\newtheorem{cor}[thm]{Corollary}
\newtheorem{lem}[thm]{Lemma}
\newtheorem{prop}[thm]{Proposition}
\newtheorem{definition}{Definition}

\newcommand{\bt}[2][]{\begin{thm}\label{#2}{\bf #1}\it}
\newcommand{\et}{\end{thm}}
\newcommand{\bl}[2][]{\begin{lem}\label{#2}{\bf #1}\it}
\newcommand{\el}{\end{lem}}
\newcommand{\bc}[2][]{\begin{cor}\label{#2}{\bf #1}\it}
\newcommand{\ec}{\end{cor}}
\newcommand{\bp}[2][]{\begin{prop}\label{#2}{\bf #1}\it}
\newcommand{\ep}{\end{prop}}
\newcommand{\bd}[2][]{\begin{definition}\label{#2}{\bf #1}\rm}
\newcommand{\ed}{\end{definition}}

\begin{document}

\title{Condensation in randomly perturbed zero-range processes}

\author{L C G del Molino, P Chleboun\footnote{Present address: Dip. Matematica, Univ. Roma Tre, Largo S.L. Murialdo, 00146 Roma, Italy}, S Grosskinsky}

\address{Centre for Complexity Science, University of Warwick, Coventry CV4 7AL, UK}
%\address{$^2$ Mathematics Institute, University of Warwick, Coventry CV4 7AL, UK}

\begin{abstract}
The zero-range process is a stochastic interacting particle system that exhibits a condensation transition under certain conditions on the dynamics. It has recently been found that a small perturbation of a generic class of jump rates leads to a drastic change of the phase diagram and prevents condensation in an extended parameter range. We complement this study with rigorous results on a finite critical density and quenched free energy in the thermodynamic limit, as well as quantitative heuristic results for small and large noise which are supported by detailed simulation data. While our new results support the initial findings, they also shed new light on the actual (limited) relevance in large finite systems, which we discuss via fundamental diagrams obtained from exact numerics for finite systems.
\end{abstract}
%\submitto{\JPA}
\maketitle

\section{Introduction\label{sec:intro}}
The zero-range process is a stochastic lattice gas where the particles hop randomly
with an on-site interaction and the jump rates $g(n)$ depend only on the local
particle number $n$. It was introduced in \cite{spitzer70} as a mathematical model for interacting
diffusing particles, and since then has been applied in a large variety of contexts, 
often under different names, (see e.g. \cite{evansetal05} and references therein). 
The model is simple enough for the steady state to factorize, on the other hand it 
exhibits an interesting condensation transition under certain conditions. When 
the particle density exceeds a critical value $\rho_c$, the system phase separates 
into a homogeneous background with density $\rho_c$ (the fluid phase) and all the excess mass 
concentrates on a single lattice site (the condensate).

Besides spatial inhomogeneities (see e.g. \cite{krugetal96}), condensation can be caused by an effective attraction between the particles if the jump rates $g(n)$ have a decreasing tail as $n\to\infty$.
A generic class of such models with a power law decay
\be\label{rates}
g(n)\simeq 1+b/n^\gamma\quad\mbox{as }n\to\infty\ ,
\ee
with positive interaction parameters $b,\gamma$ has been introduced in \cite{evans00,drouffe98}, and condensation occurs if $0<\gamma <1$ and $b>0$, or if $\gamma =1$ and $b>2$. 
Results on homogeneous zero-range condensation have been applied to many clustering phenomena in complex systems such as network rewiring \cite{angeletal06}, traffic flow \cite{Kaup05} or shaken granular media \cite{meeretal07,toeroek05}, for a review see \cite{evansetal05}.
Using a mapping to exclusion models, zero-range condensation can also be used as a generic criterion for phase separation in driven diffusive systems with one or more particle species \cite{kafrietal02}. The condensation transition in this model is now well understood \cite{godreche03,evansetal05b}, also on a mathematically rigorous level \cite{stefan,ferrarisisko,armendarizetal09}, and many variants have been studied 
\cite{evansetal05,evansetal06,lucketal07,angeletal07,stefan2,schwarzkopfetal08}. In \cite{jeon10,jeon11} the influence of specific non-random perturbations have been studied for models with asymptotically vanishing jump rates.

The assumption of strict spatial homogeneity is not very realistic in applications to real complex systems which often exhibit disorder due to local imperfections. In \cite{stefanpaul08} a randomly perturbed version of the model (\ref{rates}) has been introduced, and it turned out that an arbitrary small perturbation has a drastic effect on the critical behaviour. Using heuristic arguments it was shown that condensation occurs only if $0<\gamma <\frac{1}{2}$, significantly changing the phase diagram of the unperturbed system. These first results only applied on finite systems and crucial questions on the distribution of the critical density and whether or not the system exhibits condensation in the thermodynamic limit remained open. In this paper, we provide rigorous results on the quenched free energy and prove the existence of a finite critical density in the thermodynamic limit. We also give accurate expansion results to compute the value of thermodynamical variables and their distributions for small and large perturbations which are supported by detailed simulation results.

The paper is organized as follows: In Section \ref{sec:defs} we define the perturbed zero-range process and introduce thermodynamic variables of interest (such as the free energy) and our numerical methods. In Section \ref{sec:rigor} we derive rigorous results in the thermodynamic limit and provide expansion results for small and large noise in Sections \ref{sec:expansion} and \ref{sec:large}. In Section \ref{sec:discussion} we conclude and discuss the relevance of the thermodynamic limit results in real finite systems using exact numerics for fundamental diagrams from recursion relations.

\section{Definitions and numerical methods\label{sec:defs}}

\subsection{The disordered zero-range process}

We consider a regular, $d$-dimensional lattice $\Lambda$ of finite 
size $|\Lambda |=L$ with periodic boundary conditions. A configuration is denoted by $(\eta_x )_{x\in\Lambda}$ 
where $\eta_x \in\{ 0,1,\ldots\}$ is the occupation number at site $x$. The 
dynamics of the zero-range process is defined in continuous time, such that with rate $g_x (\eta_x )$ site 
$x\in\Lambda$ loses a particle, which moves to a randomly chosen target 
site $y$ according to some translation invariant probability distribution $p(y-x)$. For 
example in one dimension with nearest neighbour hopping, the particle 
moves to the right with probability $p$ and to the left with $1-p$. Our results do not depend on the specific choice of $p(y-x)$ as long as it is of finite range. For simplicity of presentation, we focus on 
jump rates $g_x$ given by
\be\label{rates2}
g_x (n)=e^{\sigma\xi_x(n)+b/n^\gamma} \quad\mbox{for }n\geq 1\ ,\quad g(0)=0\ .
\ee
Here $\sigma$, $b$ and $\gamma$ are positive parameters and ${\xi_x(n)}$, $x\in \Lambda$, $n\in \N$ are independent, identically distributed random variables with
\be\label{exn}
\E \big[\xi_x(n)\big] =0\ ,\quad \E\big[ \xi_x(n)^2\big] =1\ ,\quad\mbox{and}\quad \delta:=\log \E\big[ e^{-\sigma\xi_x(n)}\big] <\infty\ .
\ee
By Jensen's inequality and strict concavity of the logarithm we have $\delta >0$.
For $\sigma =0$ the asymptotic behaviour of the jump 
rates is given by (\ref{rates}) so the present model, which has been introduced in \cite{stefanpaul08}, can be interpreted as a perturbation of the generic homogeneous model. Note that the exponential form of the jump rates and the use of standardized random variables $\xi_x (n)$ is purely for notational convenience, since jump rates have to be non-negative. In simulations, we will mostly use uniform random variables $\xi_x (n)$ as a generic case, but also discuss the simplifications in the case of Gaussians in Section \ref{sec:fadisti}. Our results hold for any sufficiently regular, generic perturbation of (\ref{rates}) which does not change the expected asymptotic behaviour of the rates, i.e. for which $\lim_{n\to\infty} \E [g_x (n)]\equiv \bar g<\infty$ is finite and the same for each lattice site $x$. Perturbations with $x$-dependent asymptotic expectations lead to spatial inhomogeneities which act as an additional, independent source of condensation. This mechanism is different from the one induced by the asymptotic decay of the jump rates which we want to focus on in this paper, and has been studied previously in simpler models (see e.g. \cite{krugetal96}). 
%In this study, however, we want to focus on perturbations of the model (\ref{rates}), where the and which can thereforewhich we comment on in Section \ref{sec:prelim} before Proposition \ref{afe}. 

The main difference to these studies on purely spatial disorder \cite{krugetal96} is that in the present model the noise also depends on the occupation number at each site. In \cite{stefanpaul08} it was shown that this feature leads to a drastic change in the phase diagram for finite systems with $L$ fixed. As opposed to the unperturbed model (\ref{rates}), the perturbed model (\ref{rates2}) exhibits condensation only for $\gamma\in (0,1/2)$. This is due to the contribution of the perturbation to the partition function, and is formulated precisely in Proposition~\ref{aslimit} in Section \ref{sec:rigor}. A particularly suitable application where such perturbations are relevant is for example the bus route model \cite{busroute}, where (\ref{rates}) describes the rates at which a bus proceeds to the next stop given distance $n$ to the previous bus. The larger the distance, the slower the rate due to more passangers queueing at the stations, which can lead to a condensation of buses (on a route with periodic boundary conditions). To test robustness of this phenomenon in realistic situations, it is very natural to assume that the actual functional behaviour of the rates is of the form (\ref{rates2}) with a small random dependence on the bus $x$.

\subsection{Stationary distributions and thermodynamic variables}

It is well known (see e.g. \cite{andjel82,evansetal05}) that the above zero-range process has
grand-canonical factorized steady states $\nu_\mu^\Lambda =\otimes_{x\in\Lambda}\nu^x_\mu$. The single-site marginals are
\be\label{marginal}
\nu^x_\mu (n)=\frac{e^{n\,\mu}}{z_x (\mu )}\, w_x (n)\quad\mbox{for }n\geq 0\ ,
\ee
where the chemical potential $\mu\in\R$ fixes the particle density, and the stationary weights $w_x$ are given by the jump rates via
\be\label{wxn}
w_x (n)=\prod_{k=1}^n g_x (k)^{-1} =\exp\Big( -\sigma S_x (n)-\beta(n)\Big)\ .
\ee
Here the contribution from the perturbation, $S_x(n) =\sum_{k=1}^n \xi_x (k)$, can be interpreted as the position of a random walk on $\R$ after $n$ steps with independent increments $\xi_x(k)$. The contribution from the interaction is $\beta(n)=\sum_{k=1}^n b/k^\gamma$, and acts as an $n$-dependent drift term. 
This holds independently of the translation invariant jump distribution $p(y-x)$ and for each realization 
of the $\xi_x (n)$, i.e. $\nu_\mu^\Lambda$ is a \textit{quenched} distribution. The single-site normalization is given by the partition function
\be\label{zxmu}
z_x (\mu )=\sum_{n=0}^\infty \exp\Big( n\,\mu -\sigma S_x (n)-\beta(n)\Big)
\ee
which is strictly increasing and convex in $\mu$. 
For convergence of $z_x(\mu)<\infty$ it is necessary that the drift term $\beta(n)$ dominates the stochastic part $\sigma S_x(n)$. This has been used in \cite{stefanpaul08} to predict the phase diagram of the model, which we review in Section \ref{sec:rigor} together with our new results.

We denote
\be
f_x(\mu)=\log z_x(\mu)\ ,
\ee
and the local density can be calculated as the derivative
\be\label{rhox}
\rho_x (\mu ):=\langle\eta_x \rangle_{\nu_\mu} =\frac{\partial f_x (\mu )}{\partial\mu}\ ,
\ee
and it is a strictly increasing function of $\mu$. Here $\langle\cdot\rangle_{\nu_\mu}$ denotes an average over the quenched distribution $\nu_\mu$, and $z_x$, $\rho_x$ are still random variables w.r.t.\ the perturbation. Averages w.r.t.\ realizations of the $\xi_x (n)$ are denoted by $\E$ and determine the behaviour of the system in the thermodynamic limit. As usual for disordered systems, we consider two quantities. 
The annealed free energy is defined as
\be\label{annealed}
f_A (\mu ):=\log \E\big[ z_x (\mu )\big]\ ,
\ee
and it can be interpreted as the free energy of a site with an \textit{average} perturbation. 
Using (\ref{exn}) it can be rewritten as a shift of the homogeneous free energy
\be\label{eq:f_A}
f_A (\mu)=\log \sum_{n=0}^\infty \exp\big( n\mu + n\delta -\beta(n)\big) =f(\mu + \delta)\ ,
\ee
where $f(\mu)$ and $z(\mu)$ (and in general all quantities without subscript) refer to the homogeneous 
system with $\sigma=0$.
Given that $z(\mu)$ is monotone increasing in $\mu$, $f(\mu)$ is as well, and hence $f_A(\mu)>f(\mu)$ for all $\mu$ and all $\sigma >0$. On the other hand, the quenched free energy
\be\label{quenched}
f_Q (\mu ):=\E\big[ \log z_x (\mu )\big] =\E\big[ f_x (\mu )\big]\ ,
\ee
is the average of $f_x$ and is the physically relevant quantity in the thermodynamic limit.
In general, by Jensen's inequality and concavity of the logarithm the annealed free energy $f_A (\mu )\geq f_Q (\mu )$ provides an upper bound.

As is clear from the tail behaviour of (\ref{zxmu}) (see Section \ref{sec:rigor} for a precise statement), the grand-canonical partition function $z_x(\mu)$ does not exist for $\mu>0$, and evaluating the above quantities at $\mu=0$ yields the critical values of thermodynamic system parameters \cite{stefanpaul08}. In particular, the (average) critical density is
\be
\rho_c :=\E\big[ \rho_x (0)\big]\in [0,\infty ]\ ,
\ee
which we prove to be finite if and only if $\gamma\in (0,1/2)$ in Theorem \ref{qfe}. This is the main rigorous result of this paper.

\subsection{Numerics}

Using (\ref{eq:f_A}), the annealed free energy can be computed to arbitrary precision from the unperturbed systems. The issue is to generate reliable numerical results for quenched quantities under the presence of disorder.
Convergence of $z_x(\mu)$ 
(and hence of $f_x(\mu)$) is assured by the analytical results shown in Section \ref{sec:rigor}, however, at the critical point the effect of the perturbation is maximal and convergence 
can be very slow. Therefore it is useful to define and analyze \textit{truncated} quantities
\be
z_x^N (\mu ):=\sum_{n=0}^N \exp\Big( n\,\mu -\sigma S_x (n)-\beta(n)\Big)\ ,
\ee
and, analogously, $f_x^N$ and $\rho_x^N$. By definition,
\bes
f_x^N(\mu)\stackrel{N\to\infty}{\longrightarrow} f_x(\mu)\quad a.s.\quad\mbox{(almost surely)}\ ,
\ees
i.e. convergence holds for all \textit{typical} realizations. But the speed of convergence is random and depends on $x$. As usual, we will use empirical averages on finite systems with $|\Lambda |=L$
\be\label{fqest}
\ebo_L \big[ f_x^N (\mu )\big] :=\frac1L\sum_{x\in\Lambda} f_x^N (\mu )
\ee
to estimate $\E\big[ f_x^N (\mu )\big]$. Provided that the latter is finite, the law of large numbers implies convergence of $\ebo_L [f_x^N ]$ as $L\to\infty$ for all $N\geq 0$. Furthermore, since $f_x^N$ and $\E [f_x^N ]$ are both monotone increasing in $N$, we have
\bes
f_Q(\mu)=\lim_{N\to\infty \atop L \to \infty}\ebo_L \big[ f_x^N(\mu)\big] \ ,
\ees
and the limits for $N$ and $L$ commute.

\begin{figure}
\centering
\includegraphics[width=0.48\textwidth]{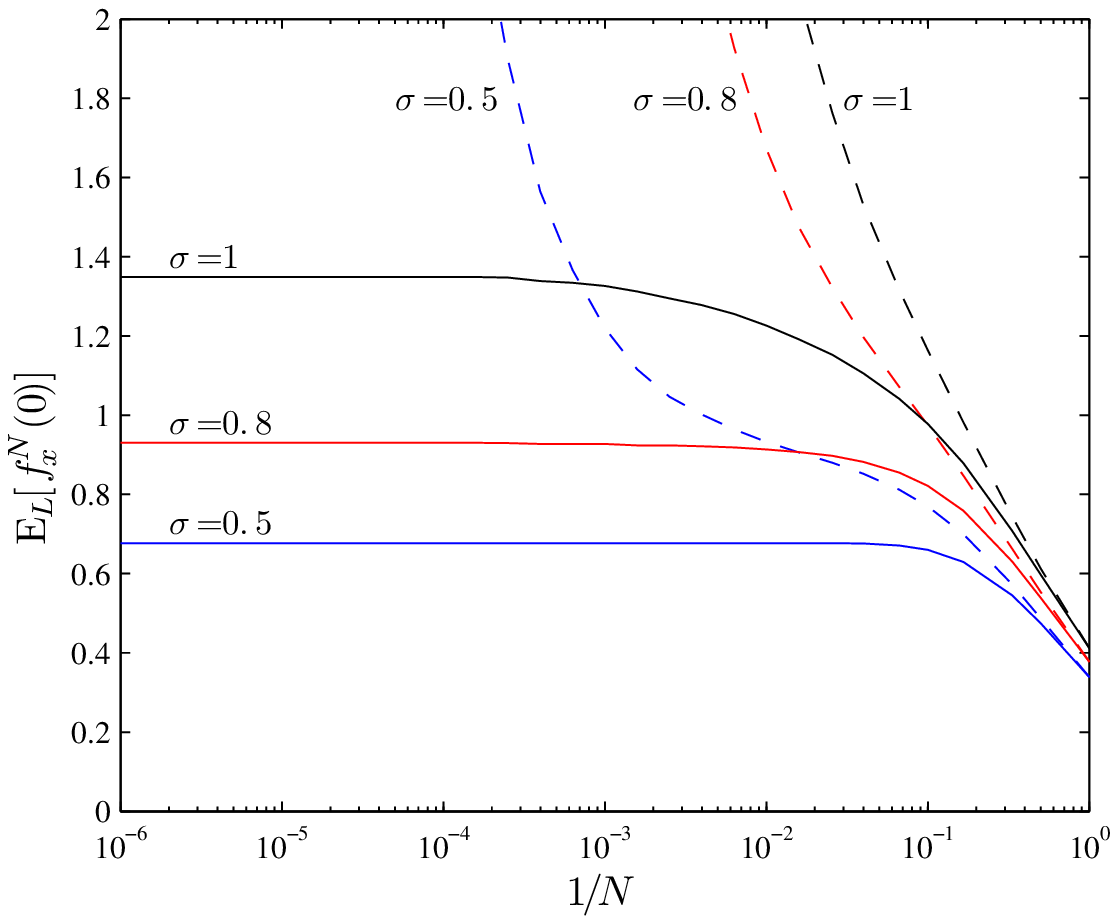}
\hfill
\includegraphics[width=0.48\textwidth]{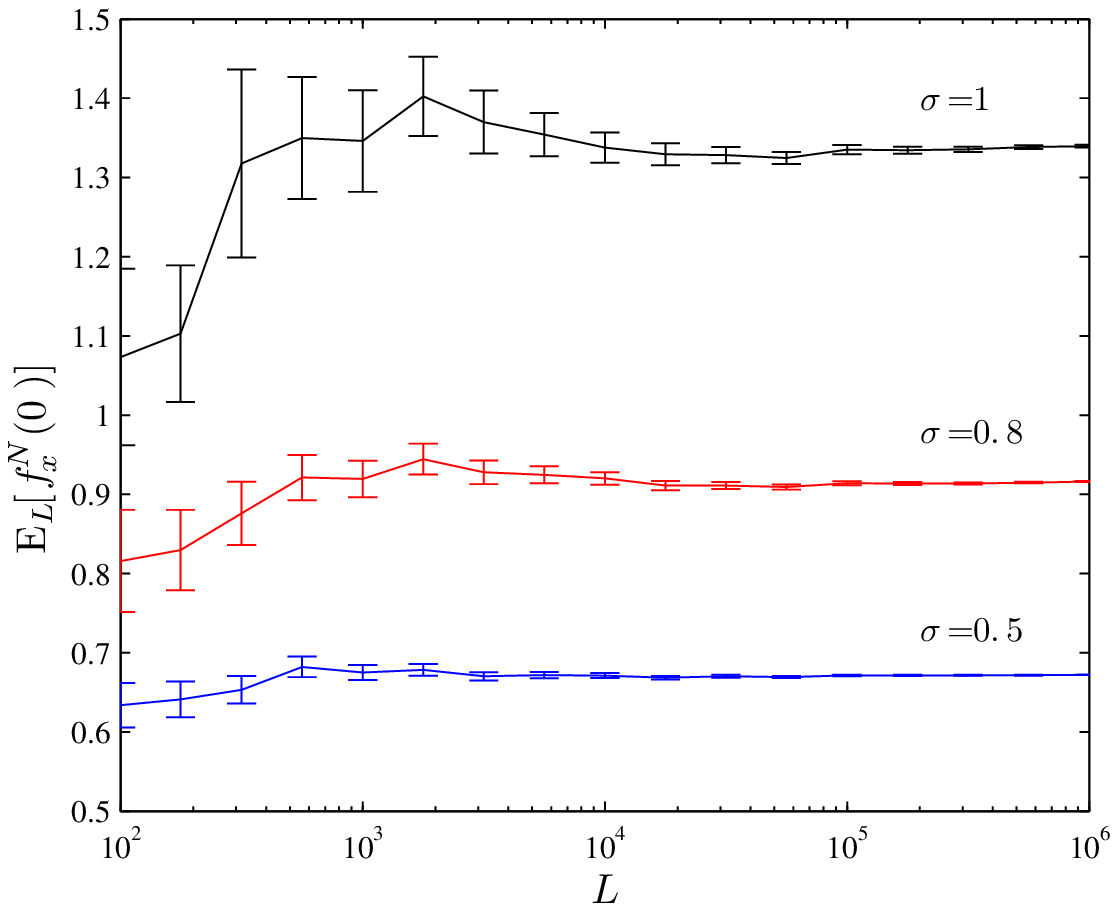}
\caption{\label{fig:convergence}Asymptotic behaviour of $\ebo_L \big[ f_x^N(0)\big]$ for $b=1$ and several values of $\sigma$ with uniform distribution of $\xi_x (n)$. (left) Fixed $L=10^4$ with increasing $N$ plotted against $1/N$. As expected, $\ebo_L \big[ f_x^N(0)\big]$ diverges for $\gamma =0.6>1/2$ (dashed lines), and for $\gamma =0.4<1/2$ (full lines) it increases monotonically to an upper bound which is the numerical estimate of $f_Q(0)$. Note that for $\gamma<1/2$ the smaller $\sigma$, the faster the convergence. (right) Fixed $N=10^5$ with increasing $L$ for $\gamma=0.4$, the error bars represent the standard error of the mean. For small $L$, rare large values dominate the empirical average at the critical point, and have to be compensated choosing $L$ large enough.
}
\end{figure}

We find that for generic values $b\approx 1$ and $\gamma\leq 0.4$ we get reliable estimates of $f_Q (\mu )$ for all $\mu\leq 0$ using
\be\label{lnparas}
N=10^5 \quad\mbox{and}\quad L=10^4 \ ,
\ee
provided perturbation strengths are $\sigma\leq 1$. This is illustrated in Fig.~\ref{fig:convergence} for $\mu =0$, where it is shown that approximations become largely independent of the truncation parameter $N$ and the number of samples $L$ above those values. We use these parameters and a uniform distribution of the $\xi_x (n)$ in all simulations in the paper, except explicitly stated otherwise.
Analogous estimates hold for $\rho_c$ and $\E\big[\rho_x (\mu )\big]$ with $\sigma\leq 0.8$, since convergence turns out to be slower for this observable (not shown).

In general, convergence becomes slower as $\sigma$ increases. A rough estimate for the minimal truncation parameter $N$ with $\mu=0$ is the point where the drift contribution $\beta(n)$ in (\ref{zxmu}) exceeds the random walk contribution $S_x(n)$ and thus successive terms of the sum start decreasing. For large $n$,
\bes
\beta(n)\sim \frac{b}{1-\gamma} n^{1-\gamma}\quad\mbox{and}\quad \big|\sigma S_x(n)\big|\sim \sigma\sqrt{n}\ , 
\ees
which leads to $N\sim \sigma^{\frac{2}{1-2\gamma}}$ (cf.\ also Section \ref{sec:large}). This can grow very fast with $\sigma$, in particular for $\gamma$ close to $1/2$, and numerical results for large perturbations are computationally expensive.

\section{Rigourous results in the thermodynamic limit\label{sec:rigor}}

\subsection{Preliminary results\label{sec:prelim}}
Before we state our main new result we summarize some simple facts for completeness, some of which have already been used in \cite{stefanpaul08}.

\bp{aslimit}
Let $\sigma >0$. $z_x(\mu)$ and $\rho_x(\mu)$ are $a.s.$ smooth functions of $\mu$ for $\mu <0$. 
For $\gamma\in [1/2,1]$ and all $b>0$ we have
\be\label{asnocond}
z_x (\mu )<\infty\ \ a.s.\ \Leftrightarrow\ \mu <0\quad\mbox{and}\quad z_x (\mu )\to\infty\ \ a.s.\quad\mbox{as }\mu\nearrow 0\ .
\ee
For $\gamma\in (0,1/2)$ and all $b>0$ we have
\be\label{ascond}
z_x (\mu )<\infty\ \ a.s.\ \Leftrightarrow\ \mu \leq 0\quad\mbox{and}\quad z_x (\mu )\to z_x (0)<\infty\ \ a.s.\quad\mbox{as }\mu\nearrow 0\ .
\ee
The same statements hold for $\rho_x (\mu )$ and all higher moments, in particular, for the site dependent critical density
\be
\rho_x (0)<\infty\ \ a.s.\quad\Leftrightarrow\quad \gamma\in (0,1/2)\ .
\ee
\ep

\noindent\textbf{Proof.} The law of the iterated logarithm (see e.g. \cite{kallenberg}) implies for the random walk part of (\ref{zxmu})
\be\label{lil}
\limsup_{n\to\infty} \frac{|\sigma S_x(n)|}{\sqrt{n\log\log n}}=\sigma\sqrt{2}\quad a.s.\ .
\ee
Furthermore, we have
\be\label{detpart}
\beta(n) =\sum_{k=1}^n \frac{b}{k^\gamma} \simeq \frac{b}{1-\gamma}\, n^{1-\gamma}\quad\mbox{as }n\to\infty\ ,
\ee
and thus\quad $\beta(n) \gg \big|\sigma S_x(n)\big|\ \ a.s.$\quad as $n\to\infty$ if and only if $\gamma\in (0,1/2)$. This implies
\be\label{asconv}
-\sigma S_x(n) -c \beta(n) \to -\infty\ \ a.s.\quad\mbox{and}\quad n^q e^{-\sigma S_x(n) -c \beta(n)} \to 0\ \ a.s.
\ee
for all $c>0$, $q>0$. In particular,\quad $w_x (n)\to 0\ \ a.s.$\quad as $n\to\infty$, and convergence is fast enough to bound the sum $z_x (0)=\sum_{n\geq 0} w_x (n)$. (\ref{asconv}) implies that
\bes
\P\big( w_x (n) >e^{-\beta(n)/2}\mbox{ for only finitely many }n\big) =1\ ,
\ees
and therefore there exists a $C>0$ such that
\be\label{zxfin}
z_x (0)\leq C\sum_{n\geq 0} e^{-\beta(n)/2} <\infty\quad a.s.\ .
\ee
The same holds for higher moments, e.g. for the density we have
\bes
\P \big( n\, w_x (n) >e^{-\beta(n)} \mbox{ for only finitely many }n\big) =1\ ,
\ees
which, together with (\ref{zxfin}), implies that $\rho_x (0)<\infty\ \ a.s.$\ .
If $\gamma \in [1/2,1)$ this argument only holds for $\mu <0$. For $\mu =0$ and $\gamma \in [1/2,1)$ (\ref{lil}) and (\ref{detpart}) imply that the random walk part dominates $w_x (n)$ for large $n$. Therefore
\bes
\P\big( w_x (n) \geq 1\mbox{ for infinitely many }n\big) =1\ ,
\ees
and $z_x (0)=\infty\ \ a.s.$\ .\hfill $\Box$\\

On fixed finite systems with lattice $\Lambda$ this directly implies that for $\gamma\in (0,1/2)$
\be\label{rhocfin}
\rho_c (\Lambda ):=\frac1L \sum_{x\in\Lambda} \rho_x (0)<\infty\quad a.s.\ ,
\ee
i.e. there exists a finite critical density, which depends on the realization of the perturbation. In correspondence to previous results such as \cite{ferrarisisko} the process is then expected to exhibit condensation when the number of particles diverges, which can be formulated in terms of canonical distributions and the equivalence of ensembles (cf.\ Section \ref{sec:discussion}). In the following we focus on properties of the grand canonical distributions in the thermodynamic limit.
% Note that for perturbations of (\ref{rates}) which lead to different expected 

\bp{afe}
The annealed free energy (\ref{annealed}) is $f_A (\mu )=f(\mu +\delta)\in [0,\infty ]$, given in terms of the unperturbed model for all $\mu\in\R$. For all $\gamma\in (0,1)$ and for $\gamma =1$ and $b>1$ we have
\be
f_A (\mu )<\infty\quad\Leftrightarrow\quad \mu\leq -\delta\ .
\ee
\ep

\noindent\textbf{Proof.} By direct computation we have
\be\label{facompu}
\E\big[ z_x (\mu )\big] =\sum_{n=0}^\infty \E\big[ w_x (n)\big] =\sum_{n=0}^\infty \E\big[ e^{-\sigma\xi_x (1)}\big]^n e^{-\beta(n) +\mu n} =z(\mu +\delta )
\ee
by monotone convergence, since the terms in the sum are positive. The rest follows immediately from the well-known properties of $f(\mu )=\log z(\mu )$ for the unperturbed model (see e.g. \cite{evans00,stefanpaul08}).\hfill $\Box$\\

This implies in particular, that even though $z_x (0)<\infty\ a.s.$ for $\gamma\in (0,1/2)$ and it is a well defined random variable, we have $\E\big[ z_x (0)\big] =\infty$. However, $\log z_x (0)$, which is the thermodynamically relevant quantity, has finite expectation $f_Q (0)$ in that case, as we will see in the next subsection.

\subsection{Main results}

\bt{qfe}
For the quenched free energy (\ref{quenched}) we have
\be\label{bounds}
f(\mu )\leq f_Q (\mu )\leq f(\mu +\delta )=f_A (\mu )\quad\mbox{for all }\mu\in\R\ .
\ee
For $\gamma\in (0,1/2)$ we have
\be\label{cond}
f_Q (\mu )<\infty\quad\Leftrightarrow\quad \mu\leq 0\quad\mbox{and}\quad\rho_c =\E\big[\rho_x (0)\big] <\infty\ .
\ee
For $\gamma\in [1/2,1]$ we have
\be\label{nocond}
f_Q (\mu )<\infty\quad\Leftrightarrow\quad \mu <0\quad\mbox{and}\quad f_Q (\mu )\to\infty\quad\mbox{as }\mu\nearrow 0\ ,
\ee
and $\rho_c =\infty$.
\et

\begin{figure}
\centering
\includegraphics[width=0.6\textwidth]{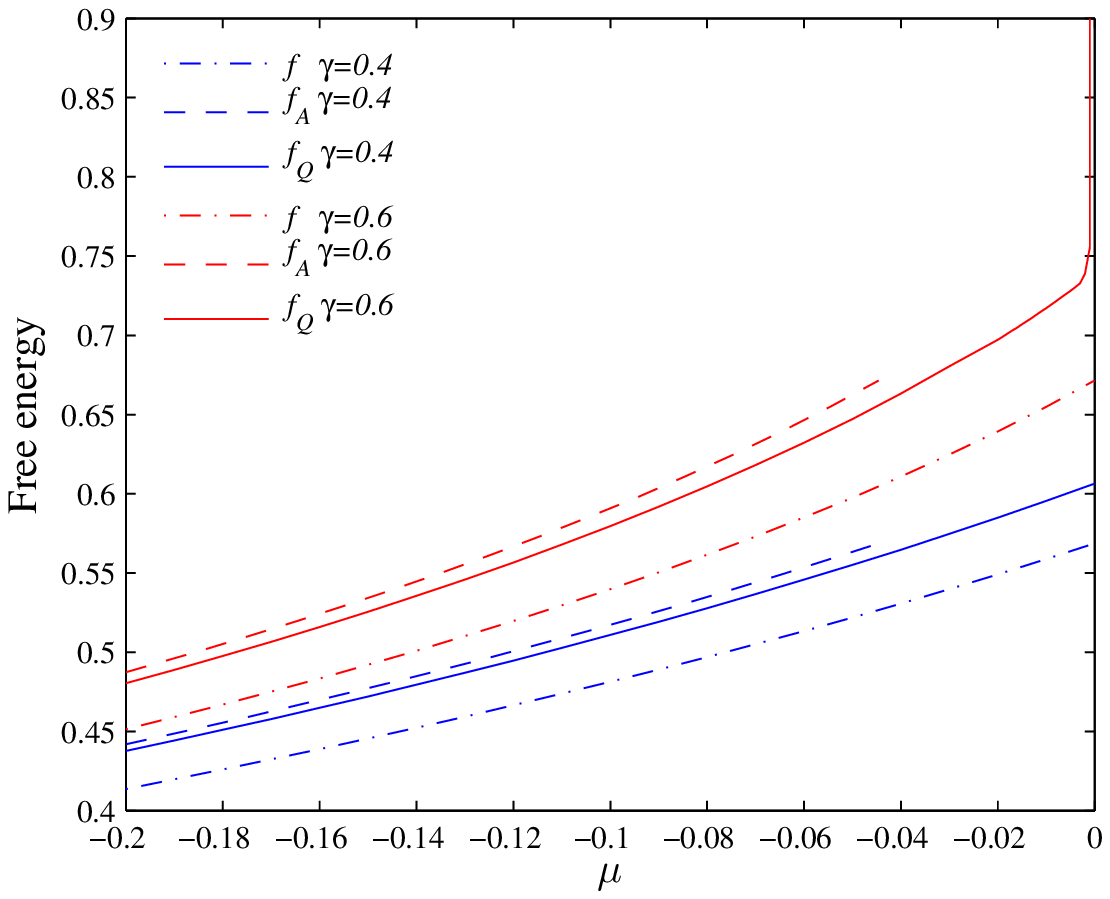}
\caption{\label{fig:freeenergy}
Upper and lower bounds (\ref{bounds}) for the quenched free energy $f_Q$ (full line) given by the annealed and unperturbed free energies $f_A $ (dashed lines) and $f$ (dash-dotted lines), respectively. Parameters are $b=1$, $\sigma =0.3$ (implying $\delta =0.0446$) for uniformly distributed noise $\xi_x (n)$, and $\gamma =0.4$ and $0.6$. $f_Q$ diverges for $\gamma =0.6$ only very close to $\mu =0$ (cf.\ also Fig.~\ref{fig:convergence}), which cannot be inferred from the behaviour of upper or lower bounds. Data for $f_Q$ are simulated according to (\ref{fqest}) and (\ref{lnparas}).
}
\end{figure}

\noindent By the definition of the critical density for finite systems (\ref{rhocfin}), and the fact that the summands are independent positive random variables, we have a strong law of large numbers
\be\label{lln}
\rho_c (\Lambda )\to \rho_c \quad a.s.\quad\mbox{as }L\to\infty\ ,
\ee
which covers condensation with $\rho_c <\infty$ for $\gamma\in (0,1/2)$, as well as the case $\rho_c =\infty$ for $\gamma\in [1/2,1]$. Fig.~\ref{fig:freeenergy} illustrates the bounds (\ref{bounds}) on the quenched free energy, which are rather accurate for small perturbation strength $\sigma$ away from the critical point, but do not contain any information on whether the system condenses or not.\\

\noindent\textbf{Proof of Theorem 3.} The upper bound in (\ref{bounds}) follows immediately from Jensen's inequality and Prop.~\ref{afe}. The lower bound follows from (\ref{lbound}) in Section \ref{sec:expansion}.

Let $\gamma\in (0,1/2)$. In the following we prove $f_Q (\mu )<\infty$ for $\mu =0$, which implies the same for all $\mu <0$ by monotonicity. Write $z_x^N(0) =\sum_{n=0}^N w_x (n)$. Then $z_x^0(0) =1$ and $z_x^N(0) >1\ a.s.$. Let $\Delta_N :=\E\big[ f_x^N(0) -f_x^{N-1}(0)\big]$. Then
\bea\label{deltan}
\Delta_{N+1} &=&\E\big[ f_x^{N+1}(0) - f_x^N(0)\big] =\E\big[\log (z_x^{N+1}(0) /z_x^N(0) )\big] =\nonumber\\
&=&\E\Big[\log \Big( 1+\frac{w_x (N+1)}{z_x^N(0)}\Big)\Big]\leq \E\Big[\frac{w_x (N+1)}{z_x^N(0)}\Big]\ ,
\eea
since $\log (1+u)\leq u$ for $u>0$. With $w_x (N+1) =w_x (N)e^{-\sigma\xi_x (N+1) -b/(N+1)^\gamma}$ and by independence of $\xi_x (N+1)$ we get
\be\label{deltan2}
\E\Big[\frac{w_x (N+1)}{z_x^N(0)}\Big] =e^{\delta -b/(N+1)^\gamma }\,\E\Big[ \frac{w_x (N)}{z_x^N(0)}\Big]\ .
\ee
Now, $w_x (N)\leq z_x^N(0) \ a.s.$ and we can estimate
\be\label{estimate}
\E\Big[ \frac{w_x (N)}{z_x^N(0)}\Big] \leq 1\times\P \Big(\frac{w_x (N)}{z_x^N(0)} >1/N^2\Big) +1/N^2\ .
\ee
Since $z_x^N(0) \geq 1\ a.s.$ we have, using the asymptotic form of $\beta (n)$ given in (\ref{detpart}),
$$
\P \Big(\frac{w_x (N)}{z_x^N(0)} >1/N^2\Big)\leq \P\big( w_x (N)>1/N^2 \big) \leq\P \Big( {-}\sigma S_x (N)>\frac12\,\frac{b}{1-\gamma}\, N^{1-\gamma} -2\log N\Big)
$$
for all $N$ large enough. Since $\gamma\in (0,1/2)$ this is bounded above by $\exp\big[ -N^{1-2\gamma}\big(\frac{b}{1-\gamma}\big)^2\big(\frac{1}{8\sigma}\big)\big]$, using standard estimates from moderate large deviations (see e.g. \cite{paul,AGL} and \cite{moderatelargedev} for a general reference). Thus with (\ref{deltan}) and (\ref{deltan2}) we have $\sum_{k=1}^\infty \Delta_k <\infty$ and
\be
f_Q (0)=\lim_{N\to\infty} \E\big[ f_x^N(0)\big] =\lim_{N\to\infty} \sum_{k=1}^N \Delta_k <\infty\ .
\ee
The second statement, $\rho_c <\infty$, can be shown very similarly. Let
\be
\rho_x^N(0) :=\frac{1}{z_x (0)}\,\sum_{n=0}^N nw_x (n) <\infty
\ee
for all $N>0$, which is well defined since $1/z_x (0)\in (0,1]\ \ a.s.$. With $\Delta_N =\rho_x^N(0) -\rho_x^{N-1}(0)$ we get
\bes
\Delta_{N} =\frac{Nw_x (N+1)}{z_x (0)}<N\quad a.s.\ .
\ees
Analogous to (\ref{estimate}) this leads to
\bes
\E [\Delta_{N}] \leq N\,\P \Big(\frac{w_x (N)}{z_x (0)} >1/N^3\Big) +1/N^2\ ,
\ees
and since $z_x (0)>1\ a.s.$ we have
$$
\P \Big(\frac{w_x (N)}{z_x (0)} >1/N^3\Big)\leq \P\big( w_x (N)>1/N^3 \big) \leq\P \Big( {-}\sigma S_x (N)>\frac12\,\frac{b}{1-\gamma}\, N^{1-\gamma} -3\log N\Big)
$$
for all $N$ large enough. This is again bounded by $\exp\big[ -N^{1-2\gamma}\big(\frac{b}{1-\gamma}\big)^2\big(\frac{1}{8\sigma}\big)\big]$ since the leading order term is unchanged. The rest follows analogously.

For $\gamma\in [1/2,1]$, on the other hand, (\ref{nocond}) follows immediately from the almost sure behaviour of $z_x (\mu )$ given in (\ref{asnocond}), and for $\mu <0$ one can use simple exponential bounds analogous to the above.\hfill $\Box$\\

This result implies that for $\gamma\in (0,1/2)$ the local particle density $\rho_x (0)$ at the critical point has finite mean. The same can be shown analogously also for all higher moments of the occupation number, which are given by higher order derivatives of $f_x (0)$. We will see in the next section using a heuristic expansion, that for small noise $f_x$ can be approximated as a small perturbation of $f$ for the unperturbed system. The actual distribution of $f_x (0)$ is very hard to describe analytically or access numerically with adequate precision, since it has heavy (sub-exponential) tails as implied by the following result.

\bp{exmoments}
Let $\gamma\in (0,1/2)$ and $b,\sigma >0$. Then we have for all $\lambda >0$
\be
\E\big[ e^{\lambda f_x (0)}\big] =\infty\ ,
\ee
i.e. the distribution of $f_x (0)$ does not have exponential moments.
\ep

\noindent Before we proceed with the proof, we introduce some notation which is used again later in Section \ref{sec:fadisti}. We would like to stress the dependence on $\sigma$ of the partition function and write
\bes
z_x (\mu,\sigma)=z_x (\mu )=\sum_{n=0}^\infty e^{n\mu - \sigma S_x(n)-\beta(n)}\ .
\ees
Recalling that $z(\mu)=\sum_{n=0}^\infty e^{n\mu-\beta(n)}$ for the unperturbed system,
we can define a probability distribution $p(n),\ n\geq 0$ where the random variable $X$ takes the value $-S_x(n)$ with probability $p(n) =\frac{e^{n\mu -\beta (n)}}{z(\mu)}$. Denoting $\langle\cdot\rangle_p$ as the expectation w.r.t.\ this distribution, we can write
\be\label{eq:jensen}
z_x (\mu ,\sigma ) =z(\mu)\,\langle e^{\sigma X}\rangle_p\ .
\ee
\mbox{}\\
\noindent\textbf{Proof of Proposition \ref{exmoments}.} Writing $f_x (\mu ,\sigma )$ analogously to $z_x (\mu ,\sigma )$, we have
\bes
\E\big[ e^{\lambda f_x (0,\sigma )}\big] =\E\big[ z_x (0,\sigma )^\lambda \big] \geq\E\big[ z_x (0,\sigma )\big]^\lambda = \infty
\ees
for $\lambda\geq 1$, directly by Jensen's inequality and (\ref{facompu}). For $\lambda \in (0,1)$ we use (\ref{eq:jensen}), and writing 
%introduced in Section~\ref{sec:fadisti} below to write
%\bes
%z_x (0)=z_x (0,\sigma )=z(0)\,\langle e^{\sigma\, X}\rangle_p \ ,
%\ees
%where $X$ takes values $-S_x (n)$ with probability $p(n)=e^{-\beta (n)}/z(0)$. Writing 
$z_x$ formally as an expectation we can apply Jensen's inequality to get
\bes
z_x (0,\sigma )^\lambda =z(0)^\lambda\,\langle e^{\sigma\, X}\rangle_p^\lambda \geq z(0)^\lambda\,\langle e^{\lambda\sigma\, X}\rangle_p =z(0)^{\lambda -1} \, z_x (0,\lambda\sigma )\ .
\ees
Now, taking expectation w.r.t.~the disroder
\bes
\E\big[ z_x (0,\sigma )^\lambda\big]\geq z(0)^{\lambda -1}\E\big[ z_x (0,\lambda\sigma )\big] =\infty\ ,
\ees
which follows from (\ref{facompu}) and using that $z(0)\in (0,\infty )$ for the unperturbed system.\hfill $\Box$\\

\begin{figure}
\centering
\includegraphics[width=0.48\textwidth]{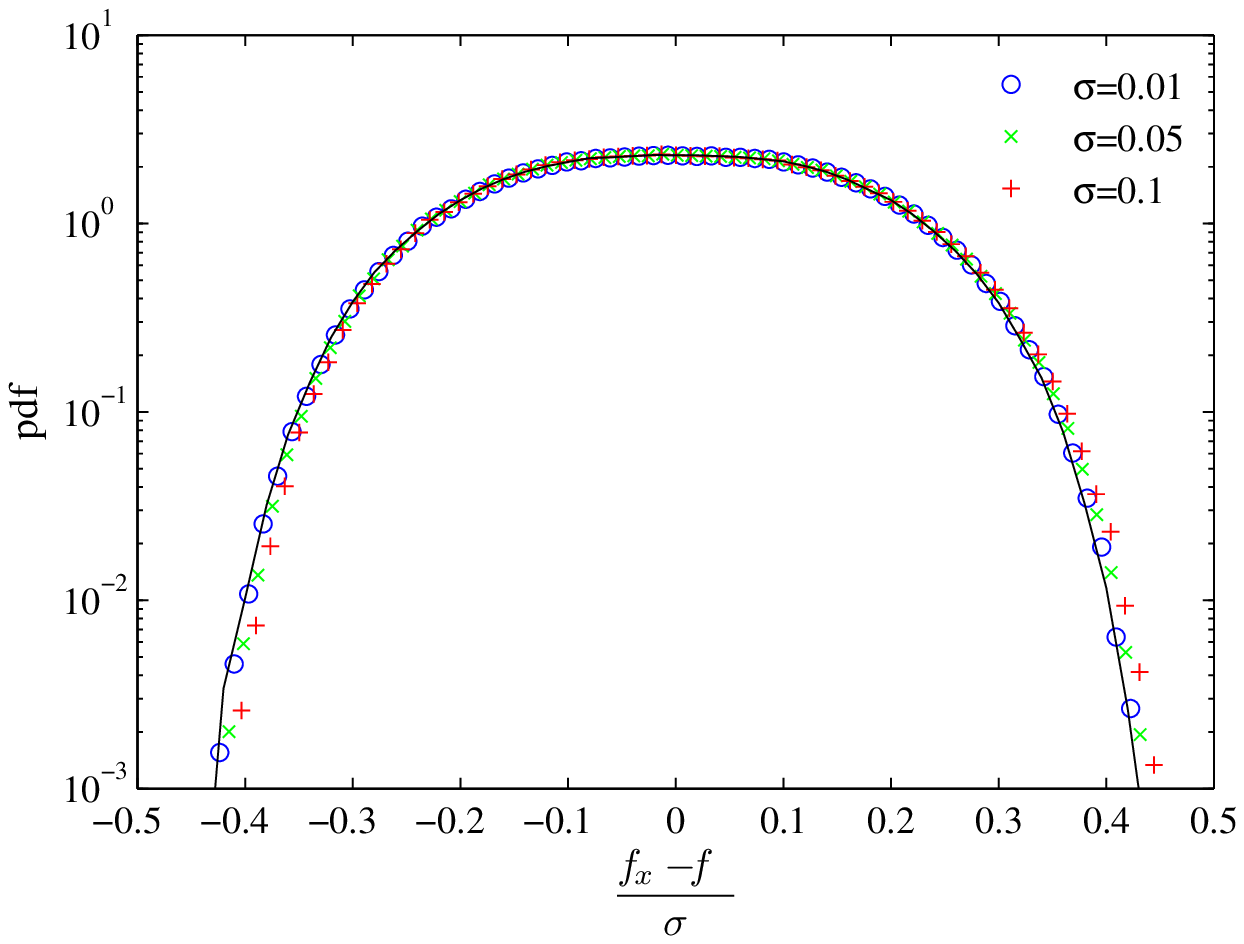}
\hfill
\includegraphics[width=0.48\textwidth]{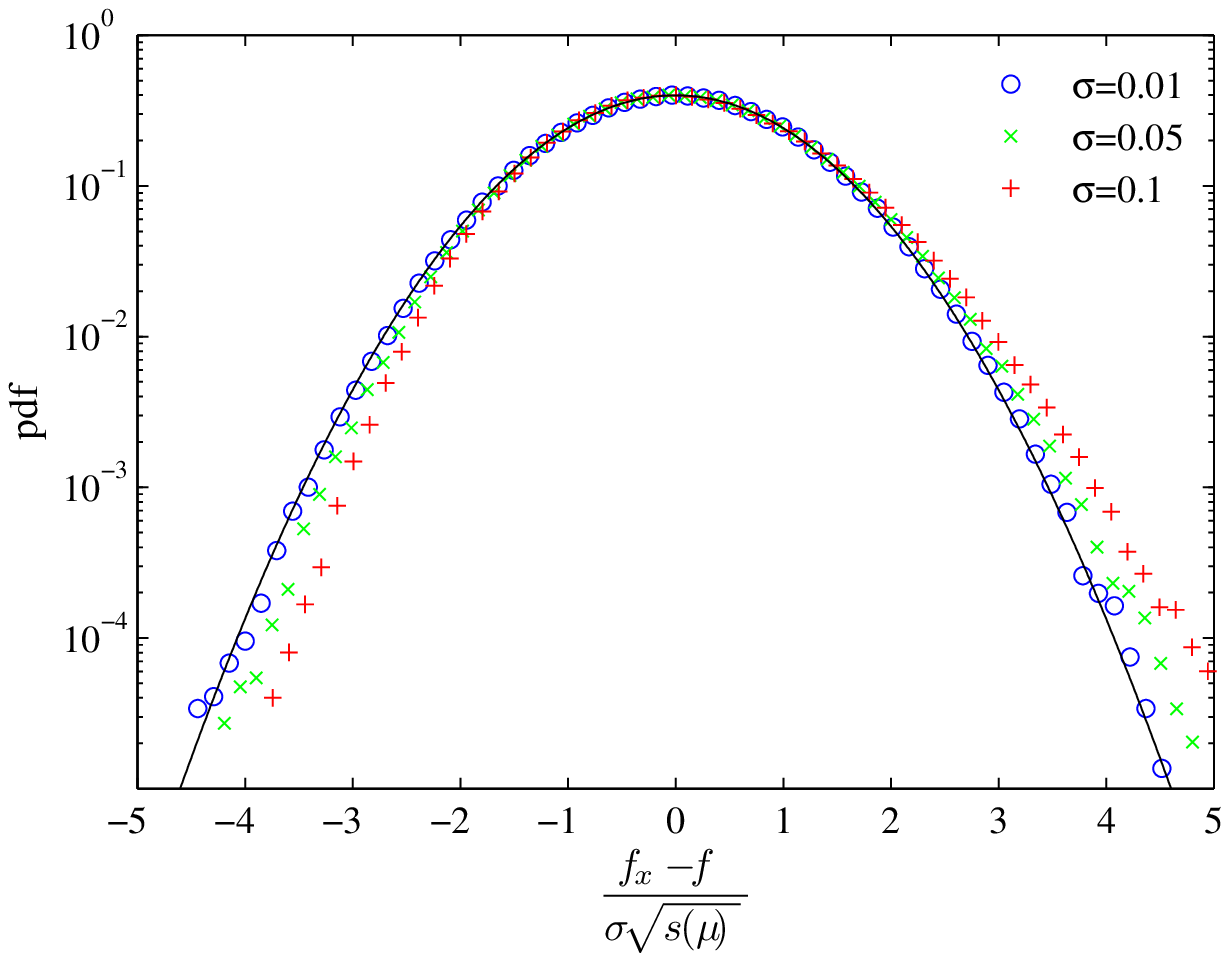}\\
{\small uniform noise}\hspace*{0.35\textwidth}{\small Gaussian noise}
\caption{\label{fig:hist}
Convergence of the rescaled probability density function of $f_x$ for vanishing $\sigma$, with $\mu=0$, $b=1$, $\gamma=0.4$ and $N=10^5$. For each value of $\sigma$ (symbols as in legend) $10^6$ samples were taken. (left) For uniform $\xi_x(n)$ the limit is non-Gaussian, and can be determined numerically from (\ref{eq:histogram}) as shown by the full line. (right) For Gaussian $\xi_x(n)$ the limit is also Gaussian and can be determined explicitly (\ref{ghisto}). The data for larger $\sigma$ is skewed towards positive values due to higher order corrections.
}
\end{figure}

\section{Expansion for small perturbation\label{sec:expansion}}

\subsection{Free energy distribution\label{sec:fadisti}}

The effects of the noise on the critical density and other thermodynamic variables are hard to quantify in general beyond the results in Section \ref{sec:rigor}, but expanding the partition function for small noise $\sigma\to 0$ leads to reasonable approximations in comparison with the unperturbed system.
% that prove to be accurate in a range of values of $\sigma$ depending on the other parameters.
With Prop.~\ref{aslimit} we have for $\gamma<1/2$, $\mu\leq 0$ and all $\sigma >0$,
% as defined in (\ref{zxmu}), we have
\bes
z_x (\mu,\sigma)=z(\mu)\,\langle e^{\sigma X}\rangle_p <\infty\quad\mbox{a.s.}\ ,
\ees
where we use the notation introduced in (\ref{eq:jensen}). Taking logarithms on both sides and using Jensen's inequality w.r.t.\ the distribution $p$ we get
\bes
\log z_x(\mu ,\sigma ) =\log z(\mu) +\log \langle e^X \rangle_p\geq \log z(\mu)+ \langle X\rangle_p\ .
\ees
Under the expectation w.r.t.\ the realizations of $S_x(n)$ the second term vanishes and we obtain a lower bound for $f_Q$,
\be\label{lbound}
f_Q(\mu) \geq f(\mu) + \E \big[ \langle X\rangle_p \big] =f(\mu)\ .
\ee

Expanding the exponential in (\ref{eq:jensen}) around $\sigma =0$ yields for a fixed realization of the noise
\be\label{expa}
z_x(\mu,\sigma)=z(\mu)\sum_{m=0}^\infty\frac{\sigma^m}{m!} \, \langle X^m \rangle_p\ .
\ee
According to Prop.~\ref{aslimit} all moments of $z_x(\mu ,\sigma)$ are finite, so $\langle X^m\rangle_p <\infty$ a.s. for all $m$. Furthermore, $\big|\langle X^m \rangle_p \big|\leq C \sum_{n=0}^\infty p_n n^{m/2}$ is almost surely bounded by the $m/2$-th moment of $p_n$, which is bounded by $m!$ for all $\mu\leq 0$. Therefore the series (\ref{expa}) converges absolutely, and $z_x(\mu ,\sigma)$ (as well as $f_x(\mu ,\sigma)$) are in fact analytic functions in $\sigma$ around $\sigma =0$ for all fixed $\mu\leq 0$.

From (\ref{expa}) we get
\bes
\log\big(z_x(\mu,\sigma)\big)=\log\big(z(\mu)\big)+\log\Big(1+\sum_{n=1}^\infty\frac{\sigma^n}{n!} \langle X^n \rangle_p \Big)\ ,
\ees
and expanding the second logarithm on the right hand side up to second order in $\sigma$ yields
\bes
f_x (\mu ,\sigma )= f(\mu) +\sigma \langle X\rangle_p +\frac{\sigma^2}{2} \Big( \langle X^2 \rangle_p -\langle X\rangle_p^2\Big) + O(\sigma^3)\ .
\ees
Using $S_x (n)=\sum_{k=1}^n \xi_x (k)$ we get for the first order term
\bes
\langle X\rangle_p ={-}\sum_{n=0}^\infty S_x(n) p(n) =0-\xi_x (1) \sum_{n=1}^\infty p(n) %\nonumber\\
-\ldots ={-}\sum_{k=1}^\infty \xi_x (k) \bar F_\mu (k)\ ,
\ees
where $\bar F_\mu (k) =\sum_{n\geq k} p(n)$ is the tail of the distribution $p(n)$. Therefore, the deviation of $f_x(\mu,\sigma)$ from the unperturbed system's free energy $f(\mu)$ scales with $\sigma$, and
\be\label{eq:histogram}
\frac{f_x (\mu ,\sigma )-f(\mu)}{\sigma} \to -\sum_{k=1}^\infty \xi_x (k) \bar F_\mu (k)\ .
\ee
This is a sum of i.i.d.\ random variables with vanishing prefactors, which depends in general on the distribution of the $\xi_x (k)$ (see Fig.~\ref{fig:hist}). In the particular case that the $\xi_x (k)$ are Gaussian this is a linear combination of independent Gaussians, and therefore,
\be\label{ghisto}
\frac{f_x (\mu ,\sigma )-f(\mu)}{\sigma\sqrt{s(\mu)}} \to N( 0,1)\quad\mbox{with}\quad s(\mu )=\sum_{k\geq 1}\bar F_\mu (k)^2 <\infty\ ,
\ee
and the fluctuations are Gaussian to leading order.

\subsection{Expected values\label{sec:exval}}

\begin{figure}
\centering
\includegraphics[width=0.48\textwidth]{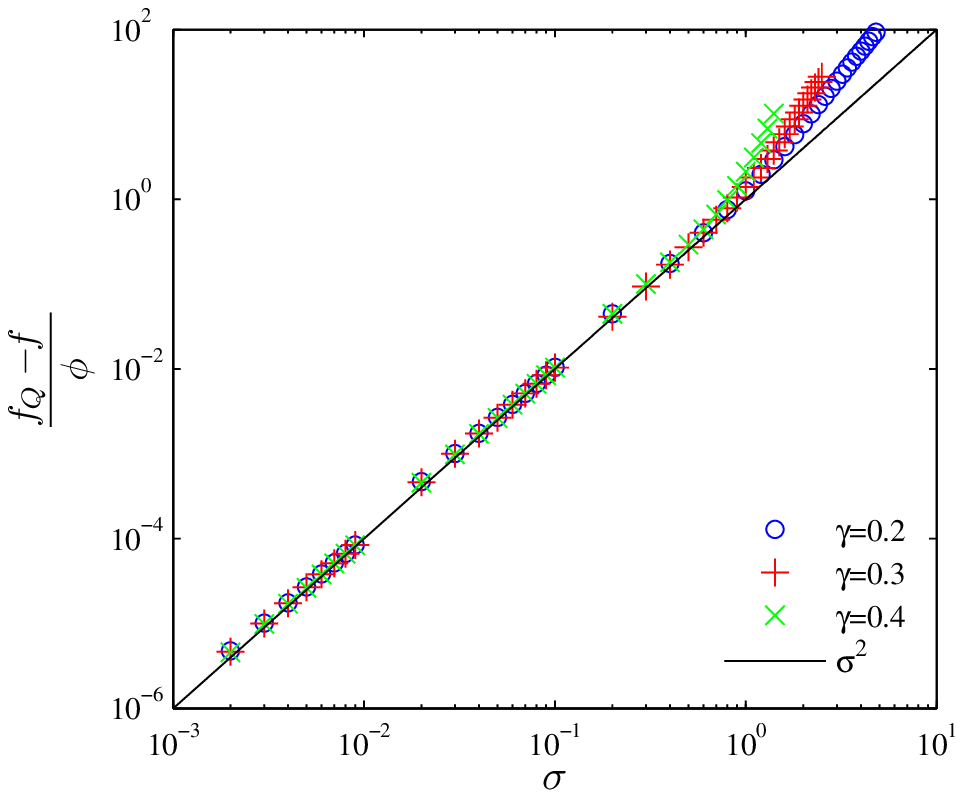}
\hfill
\includegraphics[width=0.48\textwidth]{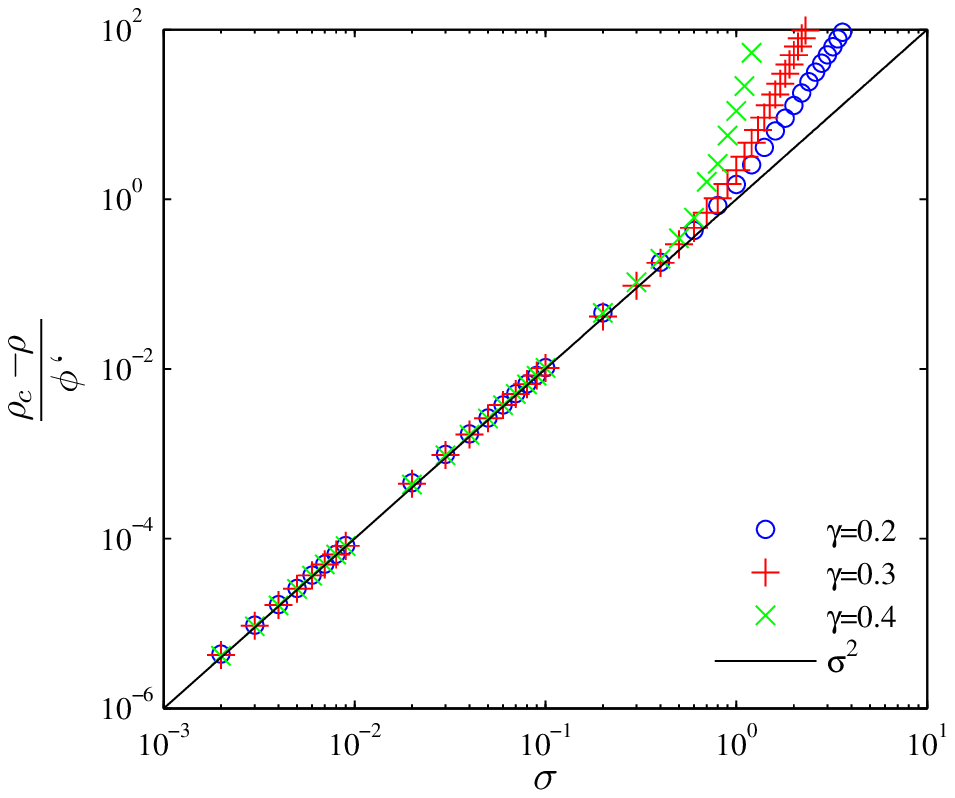}\\
{\small Free energy}\hspace*{0.35\textwidth}{\small Critical density}
\caption{\label{fig:expansion}
Convergence of the expansions of the free energy $f_Q (0,\sigma )$ (left) and the critical density $\rho_c =\E[\rho_x(0,\sigma)]$ (right) as given in (\ref{eq:fqr}) and (\ref{eq:phi}) for small disorder. Rescaled data for different values of $\gamma$ collapse well onto the predicted $\sigma^2$ behaviour (full line) for values of $\sigma$ up to approximately $0.5$. For $\sigma > 1$ clearly higher order terms become relevant. Here $N=10^5$, $L=10^5$ and errors are of the size of the symbols.}
\end{figure}

Taking expectations in the expansion (\ref{expa}) one can also estimate $f_Q=\E[ f_x ]$ and $\rho(\mu,\sigma)=\E[\rho_x(\mu ,\sigma )]$. We have $\E[S_x(n)]=0$, and also $\E [S_x(n_1) S_x(n_2) S_x(n_3)]=0$ and so on for all odd powers.
Therefore only even powers contribute and the expansion of the quenched free energy is
\be\label{eq:expfq}
f_Q (\mu ,\sigma )=\E[\log z_x(\mu,\sigma)]=f(\mu )+\frac{\sigma^2}{2}\Big(\E \left[\langle X^2 \rangle_p\right]-\E\left[\langle X\rangle_p^2\right]\Big) + O(\sigma^4)\ .
\ee
After some straightforward calculations summarized in the appendix the expectations can again be expressed in terms of the tails $\bar F_\mu (k) =\sum_{n\geq k} p(n)$. This leads to expressions for the quenched free energy and the density
\bea\label{eq:fqr}
f_Q(\mu ,\sigma )&=&f(\mu)+\sigma^2\phi(\mu)+O(\sigma^4)\nonumber\\
\E[\rho_x(\mu,\sigma)]&=&\rho(\mu)+\sigma^2\phi'(\mu)+O(\sigma^4)\ ,
\eea
where the coefficients are given by
\bea\label{eq:phi}
\phi(\mu)&=&\frac{1}{2}\sum_{n=0}^\infty \bar F_\mu (n) (1-\bar F_\mu (n))\nonumber\\
\phi'(\mu)&:=&\frac{\partial \phi(\mu)}{\partial \mu}=\frac{1}{2}\sum_{n=0}^\infty \frac{\partial \bar F_\mu(k)}{\partial \mu}(1-2\bar F_\mu(k)) \ .
\eea
Note that unlike the expressions for the distribution of $f_x(\mu)$, these results do not depend on the distribution of the $\xi_x(n)$. As is shown in Fig.~\ref{fig:expansion} the expansion coincides very well with numerical data for values of $\sigma$ even up to $0.5$.

For larger values of $\sigma$ higher orders contribute to the expansion. In a first attempt to understand these contributions one can compute the fourth order term for $\sigma^4$ of the cumulant expansion, which is given by
$$
\frac{1}{24}\E\Big[ {-}6 \langle X\rangle_p^4 {+} 12 \langle X\rangle_p^2 \langle X^2 \rangle_p {-} 3 \langle X^2\rangle_p^2 {-} 4 \langle X\rangle_p \langle X^3\rangle_p {+} \langle X^4\rangle_p\Big]\ .
$$
In order to evaluate terms of the form $\E\big[ S_x(n_1)S_x(n_2)S_x(n_3)S_x(n_4) \big]$ we can use nested conditional expectations and obtain
$$
\E\big[ S_x(n_1)S_x(n_2)S_x(n_3)S_x(n_4) \big] =3n_1^2 +n_1 (n_2 -n_1 )+n_1 (n_3 -n_2 )=2n_1^2 +n_1 n_3\ .
$$
It is difficult to find a simple formula similar to (\ref{eq:phi}) for the fourth order coefficient, but it can be computed numerically to arbitrary precision. However, including this term does not give any substantial improvement of the prediction for large $\sigma$ (not shown). Apparently, the higher order coefficients do not decay fast enough, and in particular for $\sigma >1$ the behaviour cannot be understood by looking only at the first few terms in the expansion. A different approach to tackle large disorder is presented in the next section.

\section{Large disorder\label{sec:large}}

For large $\sigma$, the disorder term can dominate the exponent in $w_x (n)$ (\ref{wxn}) up to relatively large values of $n$. For typical realizations with non-vanishing probability we can have
\be
\sigma\sqrt{2n}-\beta (n)\geq 0\quad\mbox{as long as}\quad n\leq n_* :=\Big(\frac{\sigma (1-\gamma )\sqrt{2}}{b}\Big)^{2/(1-2\gamma )}\ ,
\ee
where we have used the asymptotic form of $\beta (n)\sim \frac{b}{1-\gamma} n^{1-\gamma}$. The dominant contributions to the free energy in this case come from typical disorder realizations with large $\sigma S_x (n)\sim \sigma\sqrt{2n}$. Note that $n_*\to\infty$ with $\sigma\to\infty$, and it will serve as a large parameter in the following. Approximating the sum $z_x (0)$ (\ref{zxmu}) as an integral we get with $u=n/n_*$ replacing $\sigma S_x (n)$ by $\sigma\sqrt{2n}$,
\bea
z_x (0)&\approx &n_*\,\int_0^\infty e^{n_*^{1-\gamma} c(u)}\, du\qquad\mbox{where}\nonumber\\
c(u)&=&n_*^{\gamma -1/2} \sigma\sqrt{2}\, u-\frac{b}{1-\gamma} u^{1-\gamma}\ .
\eea
The integral is dominated by the saddle point value $u_s$ where the exponent takes its maximum, i.e.
\be
c'(u)=n_*^{\gamma -1/2}\frac{\sigma}{\sqrt{2u}} -b\, u^{-\gamma} =0\ ,
\ee
so that
\be
u_s =\bigg(\frac{\sigma n_*^{\gamma -1/2}}{\sqrt{2}b}\bigg)^{2/(1-2\gamma )} =\Big(\frac{1}{2(1-\gamma )}\Big)^{2/(1-2\gamma)} <1\ .
\ee
Expanding $c(u)\approx c(u_s )+\frac12 c''(u_s) (u-u_s )^2$ around the saddle point we get
$$
\E\big[\log z_x (0)\big]\approx \log\bigg(n_* \,\sqrt{\frac{-2\pi}{c''(u_s)\, n_*^{1\gamma}}}\, e^{n_*^{1-\gamma} c(u_s)} \bigg) = \frac12\log\frac{2\pi n_*^{1+\gamma}}{|c''(u_s)|}+C(b,\gamma )\,\sigma^{\frac{2(1-\gamma )}{1-2\gamma }}
$$
as $\sigma\to\infty$. Here $C(b,\gamma )=2^{-\frac{1-\gamma}{1-2\gamma}}\, b^{\frac{-1}{1-2\gamma}}(1-2\gamma )/(1-\gamma )$ and the lower order terms grow only logarithmically in $\sigma$ and can be ignored for large $\sigma$.

\begin{figure}
\centering
\includegraphics[width=0.48\textwidth]{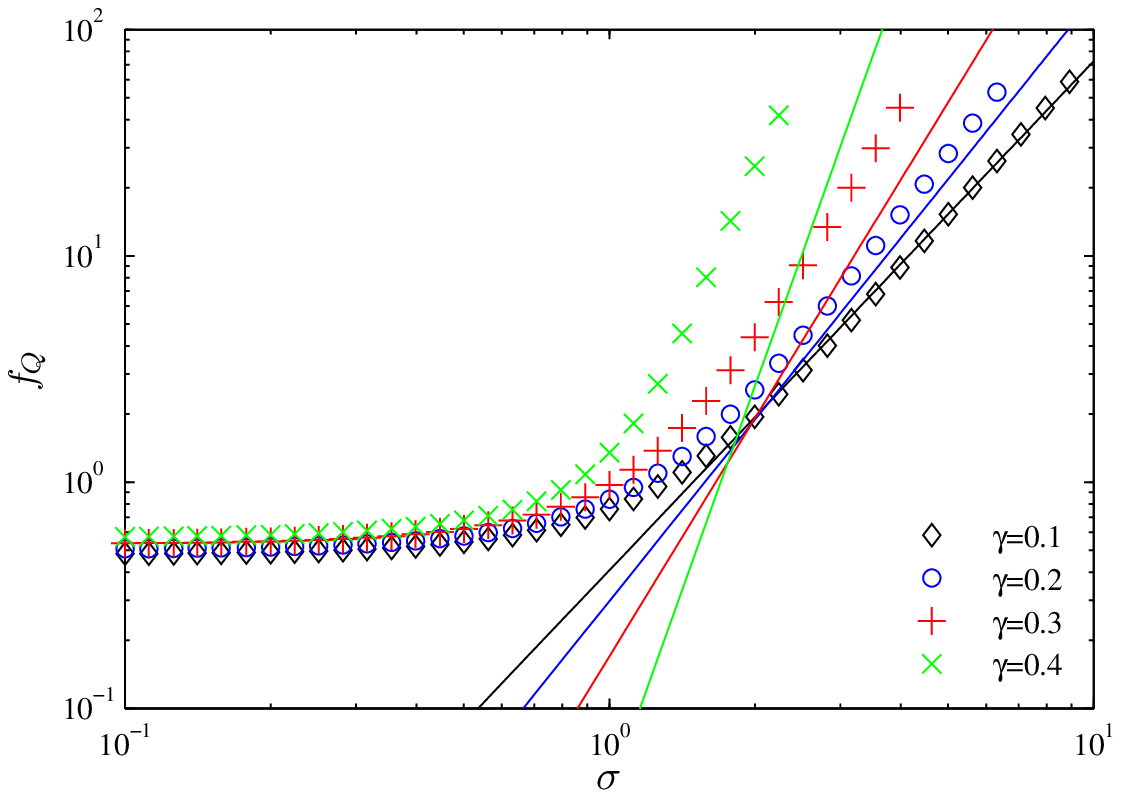}
\hfill
\includegraphics[width=0.48\textwidth]{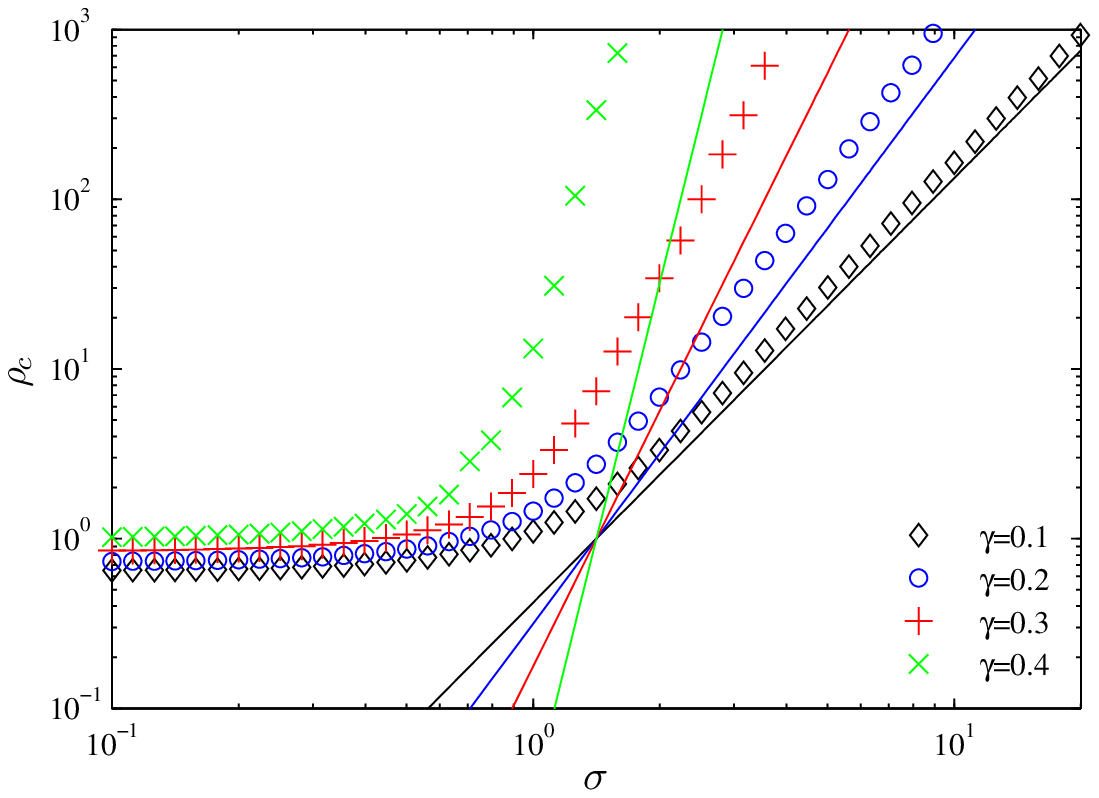}\\
{\small Free energy}\hspace*{0.35\textwidth}{\small Critical density}
\caption{\label{largesig}
Leading order behaviour of the free energy $f_Q (0,\sigma )$ (left) and the critical density $\rho_c$ (right) for large disorder. The numerical values (symbols as in legend) are expected to converge to the predictions (full lines) for $\sigma\to\infty$, and the leading order power in $\sigma$ is already well predicted also for the computationally accessible values for $\sigma\leq 10$. Here $N=10^5$, $L=10^4$ and errors are of the size of the symbols.
}
\end{figure}

A similar argument works for the critical density
\bea
\rho_c =\E \bigg[ \sum_{n=0}^\infty n\, w_x (n) \bigg/ \sum_{n=0}^\infty w_x (n)\bigg]\ ,
\eea
where both sums are dominated by the saddle point $u_s$, leading to
\bea
\rho_c \approx n_* u_s =\Big(\frac{\sigma}{\sqrt{2}b}\Big)^{2/(1-2\gamma )}\ .
\eea
Again, there are similar additive corrections as for $\log z_x (0)$ which can be ignored for large $\sigma$, and we do not write them explicitly. Both predictions are confirmed relatively well by numerical data (see Fig.~7). Since we can only go up to $\sigma =10$ due to numerical restrictions, there are still relatively large finite size corrections. But the asymptotic slope of the curves in a double-logarithmic plot corresponding to the leading order powers in $\sigma$ are well confirmed. Note that the corrections are smaller for small $\gamma$, since here the width of the Gaussian in the saddle point approximation proportional to $1/\big( n_*^{1-\gamma} |c''(u_s )|\big)$ is smaller and the integrand is concentrated more sharply around the saddle point.

\section{Discussion\label{sec:discussion}}

In this paper we have provided a fairly complete picture of the influence of a generic perturbation on the condensation transition in zero-range processes with decreasing jump rates. Our results include a rigorous analysis of the grand-canonical measures and the associated thermodynamic quantities such as the free energy and the critical density. We also provide detailed numerical data to illustrate our results, and heuristic arguments to approximately predict the behaviour of the system for small and large disorder.

\begin{figure}
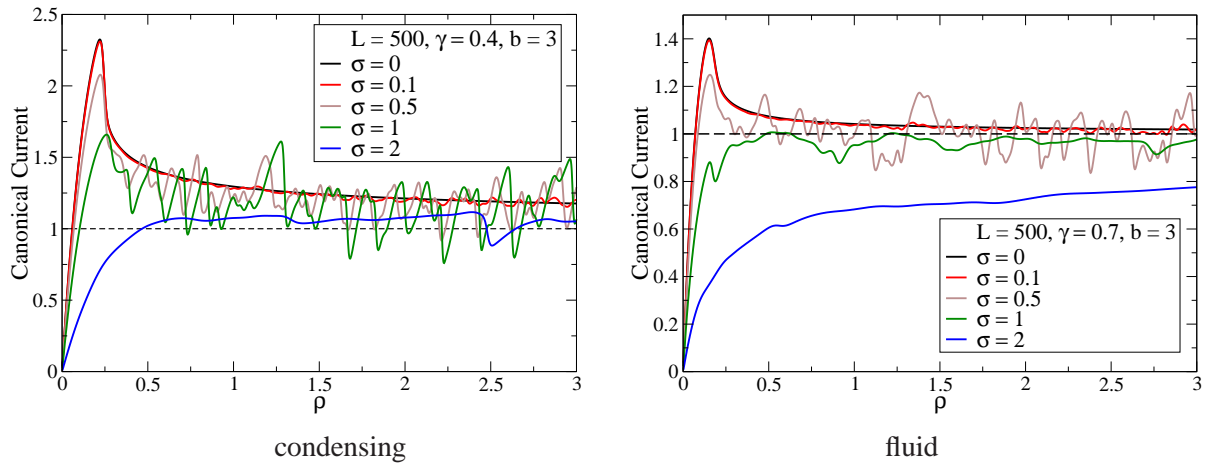

\centering
\includegraphics[width=0.48\textwidth]{currents-g04.eps}
\hfill
\includegraphics[width=0.48\textwidth]{currents-g07.eps}\\
{\small condensing}\hspace*{0.4\textwidth}{\small fluid}
\caption{\label{canonical}
Canonical current-density relation (\ref{cancurr}) for a condensing system with $\gamma =0.4$ (left) and a non-condensing system with $\gamma =0.7$ (right). For small disorder both systems appear to be condensed, and the thermodynamic limit result is only valid for large disorder or extremely large system sizes which are not accessible numerically.
}
\end{figure}

In order to understand the relevance of our results for condensation in real systems we consider the canonical stationary distribution $\pi_{\Lambda ,N}$ of a system with a fixed number of $N$ particles on the lattice $\Lambda =\{ 1,\ldots ,L\}$ with periodic boundary conditions. $\pi_{\Lambda ,N}$ can be written as a conditional distribution $\pi_{\Lambda ,N} =\nu_\mu^\Lambda \big(\, .\,\big|\sum_{x\in\Lambda} \eta_x =N\big)$, which is actually independent of $\mu$ (see e.g.~\cite{stefan}). For simplicity, we focus the discussion on totally asymmetric jumps with $p(z)=\delta_{z,1}$. In this case the average stationary current
\be\label{cancurr}
j_{\Lambda ,N} :=\big\langle g(\eta_x )\big\rangle_{\pi_{\Lambda ,N}} =\frac{Z_{\Lambda ,N-1}}{Z_{\Lambda ,N}}\ ,
\ee
is given by a ratio of canonical partition functions which can be computed exactly via the recursion relation (cf.\ e.g.\ \cite{paul,evansetal05b})
\be
Z_{\Lambda ,N} =\sum_{k=0}^N w_x (k)\, Z_{\Lambda\setminus\{ x\} ,N-k}\ .
\ee
As a reminder, for the unperturbed model (\ref{rates}) we have $z(\mu )<\infty$ if and only if $\mu\leq 0$ and $\rho_c <\infty$ as long as $\gamma\in (0,1)$ or $\gamma =1$ and $b>2$. In the thermodynamic limlit $N,L\to\infty$, $N/L\to\bar\rho$, it has been shown (see e.g. \cite{evans00,stefan}) that the canonical distributions converge to a grand canonical factorized distribution $\nu_\mu$ with spatially homogeneous marginals as given in (\ref{marginal}). For $\bar\rho\leq\rho_c$, $\mu$ is chosen to fix the density $\rho (\mu )=\bar\rho$ via relation (\ref{rhox}), and for $\bar\rho >\rho_c$, the maximal possible value $\mu=0$ (corresponding to density $\rho_c$) is chosen independently of $\bar\rho$. In the latter case the system phase separates into a homogeneous background at density $\rho_c$ with distribution $\nu_0$, and a condensate where a macroscopic fraction of all particles concentrates on a single lattice site \cite{evans00,armendarizetal09}. A particularly useful signature of the condensation transition is the behaviour of the expected current $\big\langle g(\eta_x )\big\rangle_{\nu_\mu}$ in the thermodynamic limit as a function of the density. It is monotonically increasing for densities below $\rho_c$ and becomes constant for densities above. Convergence to this limit is typically accompanied by particularly strong finite size effects which have been studied in \cite{paul}. The current shows a characteristic non-monotonic behaviour, consisting of an increasing fluid branch and a decreasing condensed branch, as can be seen in Fig.~\ref{canonical} for $\sigma =0$.

For the perturbed model we have shown in this paper that the parameter range for which condensation occurs in the thermodynamic limit changes to $\gamma\in (0,1/2)$. Nevertheless, for small noise $\sigma$ and $\gamma$ below as well as above the critical value $1/2$ the current shows the same characteristic behaviour, and the system appears to be condensing as shown in Fig.~\ref{canonical}. Only for rather large values of $\sigma\geq 1$ the system for $\gamma >1/2$ (Fig.~\ref{canonical}, right) appears fluid for all densities, as the analysis of the grand-canonical measures predicts. The large fluctuations in the current for intermediate densities result from changes in the condensate location due to the environment. The behaviour shown in Fig.~\ref{canonical} for system sizes $L=500$ is typical for all numerically accessible sizes up to $L=5000$. The thermodynamic limit results do therefore not give a good approximation of the behaviour of finite systems with moderately large system sizes, which are particularly important in many applications, such as shaken granular media \cite{meeretal07,toeroek05} or traffic flow \cite{Kaup05}.

%A particularly relevant In the spirit of mapping to latter application, the zero-range jump rates $g(n)$ correspond to exit rates of cars out of jams of length $n$, using a standard mapping to exclusion models \cite{kafrietal02,evansetal05}. Given that 

While the most interesting properties of the site-dependent free energy $f_x$, such as finite mean and sub-exponential tails, are included in our results, it would be interesting to estimate the exact tail behaviour of its distribution. This requires an understanding of the leading order contributions to the partition function which is an interesting question in itself. While properties of exponential functionals of Brownian motions with constant drift are known to great detail (see e.g.\ \cite{yor}), the form of the weights $w_x (n)$ in the present model do not allow for an exact analysis. First numerical results indicate a crossover in the behaviour, where depending on the system parameters the sum is dominated by a large number of small contributions or a small number of large contributions.

\section*{Acknowledgements}
L.C.G.M.\ was funded by the Erasmus Mundus Masters Course CSSM, and P.C.\ and S.G.\ acknowledge
support by EPSRC, grant no. EP/E501311/1.

\appendix

\section*{Appendix. Calculation of expansion coefficients}

In the following we compute the coefficients (\ref{eq:phi}) of the expansion in Section \ref{sec:exval}. Since $S_x$ has independent increments, $\E [S_x(n) S_x(k)]=\min\{n, k\}$. The first expected value in the bracket in (\ref{eq:expfq}) is
$\sum_{n=0}^\infty np(n)=\rho(\mu)$. For the second expected value we need to compute terms of the form
\bes
\sum_{n,k=0}^\infty \min\{n, k\}\, p(n) p(k) =-\sum_{n=0}^\infty n p(n)^2 +2\sum_{n=0}^\infty np(n) \bar F_\mu (n)\ ,
\ees
where $\bar F_\mu (k) =\sum_{n\geq k} p(n)$. Now use $p(n) =\bar F_\mu (n)-\bar F_\mu (n+1)$ and the trick
$$
\big(\bar F_\mu (n)-\bar F_\mu (n+1)\big)\bar F_\mu (n) =\frac12 \big( \bar F_\mu (n)^2-\bar F_\mu (n+1)^2 +\underbrace{(\bar F_\mu (n)-\bar F_\mu (n+1))^2}_{p(n)^2}\big)\ .
$$
Then summation by parts
\bes
\sum_{n=0}^\infty n\,\big( \bar F_\mu (n+1)^2-\bar F_\mu (n)^2 \big) =-\sum_{n=0}^\infty \bar F_\mu (n)^2
\ees
leads to
\bes
\sum_{n,k=0}^\infty \min\{n, k\}\, p(n) p(k) =\sum_{n=0}^\infty \bar F_\mu (n)^2 \ .
\ees
One finally obtains
$$
\E \left[ \langle X^2 \rangle_p -\langle X\rangle_p^2 \right] =\rho (\mu ,0) -\sum_{n=0}^\infty \bar F_\mu (n)^2 =\sum_{n=0}^\infty \bar F_\mu (n) \big( 1-\bar F_\mu (n)\big)\ .
$$
An estimate of the density follows by differentiation of the free energy expansion w.r.t.\ $\mu$. It is useful to compute first
\bes
\frac{\partial \bar F_\mu(k)}{\partial \mu}=\sum_{n\geq k}p(n)\left(n-\rho(\mu)\right)=\sum_{n\geq k}\bar F_\mu(n)-\rho(\mu)\bar F_\mu(k)\ .
\ees


\section*{References}

\begin{thebibliography}{30}

\bibitem{spitzer70}
Spitzer F 1970 %Interaction of Markov processes, 
\textit{Adv.\ Math.} \textbf{5} 246--290

\bibitem{evansetal05}
Evans M R and Hanney T 2005
%Nonequilibrium statistical mechanics of the zero-range  process and related models, 
\textit{J.\ Phys.\ A: Math.\ Gen.} \textbf{38} R195--R239

%\bibitem{meeretal02}
%J. Eggers, Phys. Rev. Lett. \textbf{83}(25), 5322-5325 (1999).
%D.~van der Meer, J.P.~van der Weele, D.~Lohse,
%Sudden collapse of a granular cluster. 
%Phys. Rev. Lett. \textbf{88}, 174302 (2002). F.~Coppex, M.~Droz, A.~Lipowski, Phys.~Rev.~E \textbf{66}, 011305 (2002).

\bibitem{evans96}
Evans M R 1996
%Bose-Einstein condensation in disordered exclusion models and relation to traffic flow
\textit{Europhys. Lett.} \textbf{36} 13--18

\bibitem{krugetal96}
Krug J and Ferrari P A 1996
%Phase transitions in driven diffusive systems with random rates, 
\textit{J.~Phys.~A: Math.~Gen.} \textbf{29} L465--L471

%\bibitem{benjaminietal96}
%I.~Benjamini, P.A.~Ferrari, C.~Landim, 
%Asymmetric conservative processes with random rates, 
%Stoch.~Proc.~Appl.~\textbf{61}, 181--204 (1996).
%C.~Landim, Hydrodynamic limit for space inhomogeneous one-dimensional totally asymmetric zero-range processes. Ann. Prob. \textbf{24}(2), 599-638 (1996).
%E.D. Andjel, P.A. Ferrari, H. Guiol, C. Landim: Convergence to the maximal invariant measure for a zero-range process with random rates. Stoch. Proc. Appl. \textbf{90}, 67-81 (2000).

\bibitem{evans00}
Evans M R 2000
%Phase transitions in one-dimensional nonequilibrium systems, 
\textit{Braz.\ J.\ Phys.} \textbf{30}(1) 42--57

\bibitem{drouffe98}
Drouffe J-M, Godr\'eche C and Camia F 1998
%A simple stochastic model for the dynamics of condensation, 
\textit{J.\ Phys.\ A: Math.\ Gen.} \textbf{31} L19

\bibitem{angeletal06}
%A.G. Angel, M.R. Evans, E. Levine, D. Mukamel, Phys. Rev. E \textbf{72}, 046132 (2005).
Angel A G, Hanney T and Evans M R 2006 \textit{Phys. Rev. E} \textbf{73} 016105

\bibitem{Kaup05}
Kaupuzs J, Mahnke R and Harris R J 2005
%Zero-range model of traffic flow,
\textit{Phys. Rev. E} \textbf{72}(5) 056125

\bibitem{meeretal07}
van der Meer D, van der Weele K, Reimann P and Lohse D 2007
%Compartmentalized granular gases: flux model results, 
\textit{J. Stat. Mech.: Theor. Exp.} P07021

\bibitem{toeroek05}
T\"or\"ok J 2005
%Analytic study of clustering in shaken granular material using zero-range processes. 
\textit{Physica A} \textbf{355} 374--382

%\bibitem{schlichtingetal96} 
%H. J. Schlichting and V. Nordmeier, Math. Naturwiss.
%Unterr. 49, 323 (1996) (in German).

\bibitem{kafrietal02}
Kafri Y, Levine E, Mukamel D, Sch\"utz G M and T\"or\"ok J 2002
%\newblock Criterion for phase separation in one-dimensional driven systems.
\textit{Phys.\ Rev.\ Lett.} \textbf{89}(3) 035702

\bibitem{godreche03}
Godr\'eche C 2003
%Dynamics of condensation in zero-range processes,
\textit{J.\ Phys.\ A: Math.\ Gen.} \textbf{36}(23) 6313--6328

\bibitem{evansetal05b}
Evans M R, Majumdar S N and Zia R K P 2006
%Canonical analysis of condensation in factorised steady state, 
\textit{J.\ Stat.\ Phys.} \textbf{123} 357--390

\bibitem{stefan}
Grosskinsky S, Sch\"utz G M and Spohn H 2003
%Condensation in the zero range process: stationary and dynamical properties, 
\textit{J.\ Stat.\ Phys.} \textbf{113}(3/4) 389--410

\bibitem{ferrarisisko}
Ferrari P A, Landim C and Sisko V V 2007
%Condensation for a fixed number of independent random variables, 
\textit{J.~Stat.~Phys.} \textbf{128} 1153--1158

\bibitem{armendarizetal09}
Armend{\'a}riz I and Loulakis M 2009
%Thermodynamic limit for the invariant measures in supercritical zero range processes,
\textit{Probab.\ Theory Relat.\ Fields} \textbf{145}(1) 175--188

\bibitem{evansetal06}
Evans M R, Hanney T and Majumdar S N 2006
%Interaction driven real-space condensation, 
\textit{Phys. Rev. Lett.} \textbf{97} 010602

\bibitem{lucketal07}
Luck J M and Godreche C 2007
%Structure of the stationary state of the asymmetric target process
\textit{J. Stat. Mech.: Theor. Exp.} P08005

\bibitem{angeletal07}
Angel A G, Evans M R, Levine E and Mukamel D 2007
%Criticality and Condensation in a Non-Conserving Zero Range Process, 
\textit{J.~Stat.~Mech.: Theor. Exp.} P08017

\bibitem{stefan2}
Grosskinsky S and Sch\"utz G M 2008
%Discontinuous condensation transition and nonequivalence of ensembles in a zero-range process, 
\textit{J.~Stat.~Phys.} \textbf{132}(1) 77--108

\bibitem{schwarzkopfetal08}
Schwarzkopf Y, Evans M R and Mukamel D 2008
%Zero-Range Processes with Multiple Condensates: Statics and Dynamics, 
\textit{J. Phys. A: Math. Theor.} \textbf{41} 205001

\bibitem{jeon10}
Jeon I 2010 
%Phase transition for perfect condensation and instability under the perturbations on jump rates of the zero-range process 
\textit{J.~Phys.~A:~Math.~Theor.} \textbf{43}(23) 235002

\bibitem{jeon11}
Jeon I 2011 
%Condensation in perturbed zero-range processes 
\textit{J.~Phys.~A:~Math.~Theor.} \textbf{44}(25) 255002

\bibitem{stefanpaul08}
Grosskinsky S, Chleboun P and Sch\"utz G M 2008
%Instability of condensation in the zero-range process with random interaction, 
\textit{Phys.~Rev.~E} \textbf{78}(3) 030101(R)

\bibitem{busroute}
O'Loan O J, Evans M R and Cates M E 1998
%Jamming transition in a homogeneous one-dimensional system: The bus route model
\textit{Phys.~Rev.~E.} \textbf{58}(2) 1404--1418

\bibitem{andjel82}
Andjel E 1982
%Invariant measures for the zero range process, 
\textit{Ann.\ Probability} \textbf{10}(3) 525--547

\bibitem{kallenberg}
Kallenberg O 2002 {\it Foundations of Modern Probability} 2nd edition (Springer, New York)
%L.~Breiman, Probability (1968) Addison-Wesley

\bibitem{kipnislandim}
Kipnis C and Landim C 1999
{\it Scaling Limits of Interacting Particle Systems}
%Volume 320 of Grundlehren der mathematischen Wissenschaften
(Springer, New York)

\bibitem{angeletal04}
Angel A G, Evans M R and Mukamel D 2004
%Condensation Transitions in a One-Dimensional Zero-Range Process with a Single Defect Site,
\textit{J.~Stat.~Mech.: Theor. Exp.} P04001

\bibitem{yor}
Yor M 1992
%On certain exponential functionals of real-valued Brownian motion, 
\textit{J.~Appl.~Prob.} \textbf{29} 202--208

\bibitem{paul}
Chleboun P and Grosskinsky S 2010
%Finite size effects and metastability in zero-range condensation
\textit{J.~Stat.~Phys.} \textbf{140}(5) 846--872

\bibitem{AGL}
Armend{\'a}riz I, Grosskinsky S and Loulakis M 2009
%Zero Range Condensation at Criticality,
\textit{Preprint} arXiv:0912.1793

\bibitem{moderatelargedev}
Dembo A and Zeitouni O 1998
\textit{Large Deviations Techniques and Applications}
(Springer, New York)
%Application of Mathematics, vol. 38

\end{thebibliography}
\end{document}